\documentstyle[prl,aps,epsf,floats,twocolumn]{revtex}
\begin{document}
\draft
\wideabs{
\title{Efficient algorithm for detecting unstable periodic orbits
in chaotic systems}
\author{Ruslan L. Davidchack$^1$ and Ying-Cheng Lai$^{1,2}$}
\address{1. Department of Physics and Astronomy,
         University of Kansas, Lawrence, Kansas 66045.}
\address{2. Department of Mathematics,
         University of Kansas, Lawrence, Kansas 66045.}
\date{\today}
\maketitle
\begin{abstract}

We present an efficient method for fast, complete, and accurate 
detection of unstable periodic orbits in chaotic systems.
Our method consists of a new iterative scheme and an effective
technique for selecting initial points. The iterative scheme
is based on the semi-implicit Euler method, which has both
fast and global convergence, and only a small number of initial
points is sufficient to detect all unstable periodic orbits of a
given period.  The power of our method is illustrated by
numerical examples of both two- and four-dimensional maps.

\end{abstract}
 }

It has now been a widely accepted notion that unstable periodic orbits
(UPOs) constitute the most fundamental building blocks of a chaotic 
system \cite{UPO_general}.  Theoretically, the infinite number
of UPOs embedded in a chaotic invariant set provides a skeleton of
the set, and many dynamical invariants of physical 
interest, such as the natural measure, the spectra of Lyapunov 
exponents and fractal dimensions, as well as other statistical 
averages of physical measurements, can be computed from the infinite 
set of UPOs in a fundamental way \cite{GOY:1988}.  In Hamiltonian 
systems, the quantum mechanical density of states in the semiclassical 
regime can be expressed explicitly in terms of UPOs of the 
corresponding classical dynamics \cite{Gutzwiller:1990}.
The knowledge of UPOs is also of significant experimental interest 
because they provide a way to characterize and understand the chaotic 
dynamics of the underlying system \cite{UPO_exp}.
All these call for efficient techniques for detecting UPOs in 
chaotic systems.

Systematic detection of a complete set of UPOs of high periods 
embedded in a chaotic set even in situations where the system's 
equations are known is, however, an extremely difficult problem.
A fundamental reason is that the number of UPOs grows exponentially
as the period increases at a rate given by the topological entropy
of the chaotic set. The basic requirements for a good detection
algorithm are, therefore, fast convergence and the ability to yield
complete set of UPOs \cite{BW:1989}.   

Recently, a general algorithm for detecting UPOs in chaotic systems
was proposed by Schmelcher and Diakonos (SD) \cite{SD:1997} who,
for the first time, computed UPOs of high periods for systems
such as the Ikeda-Hammel-Jones-Moloney map \cite{Ikeda:1979}.
The success of the SD method relies on a globally convergent iterative
scheme: convergence to UPOs can be achieved, in principle, from 
any initial point.  However, as we will discuss shortly, this method  
is not very efficient from the standpoint of convergence, neither 
does it provide a satisfactory test for the completeness of the 
detected UPOs.  As a matter of fact, for the 
Ikeda-Hammel-Jones-Moloney map, only 
UPOs of periods up to 13 were reported in Ref. \cite{SD:1997}, and 
one of the UPOs of period 10 was not detected.

The aim of this Letter is to present an {\it efficient} method
for detecting UPOs in general chaotic systems. Our new iterative 
scheme is based on the semi-implicit Euler method \cite{PressBook}
and has the following favorable properties: near an orbit point
it exhibits a fast convergence similar to that of the traditional
Newton-Raphson (NR) method, while away from the orbit points it
is similar to the SD method and, therefore, is
globally convergent.  Another key ingredient of our method is that
we select initial points based on the observation that 
using orbit points of UPOs of other periods to initialize the 
search for UPOs of a given period is {\it much more effective} than
using randomly selected points in the phase space or in the
attractor.  We find, in most cases, it is sufficient to use only 
orbit points of period $p-1$ in order to detect all UPOs of 
period $p$.
With such a strategy, we are able to compute UPOs for, say, the
Ikeda-Hammel-Jones-Moloney map, of periods up to 22 for a total
of over $10^6$ orbit points using roughly the same amount of 
computation required by the SD method to compute all UPOs of
periods up to 13 that have less than 6000 orbit points 
\cite{Impractical}. 
Due to its efficiency, our method allows us to 
compute UPOs in higher-dimensional systems, which we illustrate
using a four-dimensional chaotic map.

We begin by describing the NR and the SD methods. Consider an 
$N$-dimensional chaotic map: ${\bf x}_{n+1} = {\bf f}({\bf x}_n)$.
The orbit points of period $p$ can be detected as 
{\em zeros} of the following function:
\begin{equation}  \label{eq:funcg}
  {\bf g}({\bf x}) = {\bf f}^{(p)}({\bf x}) - {\bf x}\:,
\end{equation}
where ${\bf f}^{(p)}({\bf x})$ is the $p$-times iterated map of 
${\bf f}({\bf x})$. The process of finding zeros of ${\bf g}({\bf x})$
usually begins with the choice of initial point ${\bf x}_0$ followed
by the computation of successive corrections:  
${\bf x}_{\mathrm new} = {\bf x}_{\mathrm old} +\delta{\bf x}$, which
converge to the desired solution. In the NR method, 
the corrections are calculated from a set of $N$ linear equations:
\begin{equation}  \label{eq:nr}
  -{\bf J}({\bf x}) \delta{\bf x} = {\bf g}({\bf x})\:,
\end{equation}
where  ${\bf J}({\bf x}) = \partial {\bf g}/\partial {\bf x}$
is the Jacobian matrix.  The NR method has excellent 
convergence properties, approximately doubling the number of 
significant digits upon every iteration, provided that the initial
point is within the linear neighborhood of the solution.
While it is relatively easy to find suitable initial points for 
very small periods (using, for example, uniform grid, iterations of 
the map, or random number generator), the method becomes impractical 
for UPOs of high periods because the volume of the basin from which
${\bf x}_0$ can be chosen decreases exponentially as the period 
increases.  On the other hand, in the SD method, the corrections 
are determined as follows:
\begin{equation}  \label{eq:sd}
  \delta{\bf x} = \lambda {\bf C} {\bf g}({\bf x})\:,
\end{equation}
where $\lambda$ is a small positive number and ${\bf C}$ is an
$N\times N$ matrix with elements $C_{ij} \in \{0,\pm 1\}$ 
such that each row or column contains only one element that is
different from zero.  With an appropriate choice of ${\bf C}$ and
a sufficiently small value of $\lambda$ the above procedure
can find any periodic point of a chaotic system.  The main
advantage of the SD method is that the basin of attraction of each
UPO extends far beyond its linear neighborhood, so most
initial points converge to a UPO. In fact, the basins of
attraction of individual orbit points completely fill a region 
in the phase space, and any initial point in this 
region converges to an orbit point.  

Schmelcher and Diakonos tested their method by computing the
UPOs for the H\'{e}non map and other simple maps,
for which the UPOs are known from methods specific to these maps
\cite{BW:1989}. 
They also applied the method to the Ikeda-Hammel-Jones-Moloney map,  
for which no special technique for computing UPOs was previously 
available. The method appears to be particularly useful when detecting
for each period the least unstable periodic orbits \cite{DSB:1998}.
However, if the goal is to determine complete sets of UPOs of
increasingly higher periods, the SD method
becomes inefficient due to the following two reasons: (i)
the convergence rate of Eq.~(\ref{eq:sd}) is much slower than
that of the NR method, so it takes significantly more steps
to reach the desired accuracy \cite{Convergence}; and (ii)
even though the SD scheme is globally convergent, 
the basins of attraction of
individual UPOs are interwoven in a complicated manner, so it is
extremely difficult to determine which initial point converges to a
particular UPO.  Because of this difficulty, the SD method cannot
guarantee the detection of all UPOs of a given period.

To overcome the problem of slow convergence, while retaining the
global convergence property, we propose the following
iteration scheme:
\begin{equation}  \label{eq:ours}
  [{\bf 1}\beta g({\bf x}) - {\bf C}{\bf J}({\bf x})] 
  \delta{\bf x} = {\bf C}{\bf g}({\bf x})\:,
\end{equation}
where $g({\bf x}) \equiv ||{\bf g}({\bf x})|| \ge 0$ is the
length of the vector, $\beta > 0$ is an adjustable parameter, and
${\bf C}$ is the same
\begin{figure}
  \epsfxsize=8.3cm  \epsfbox{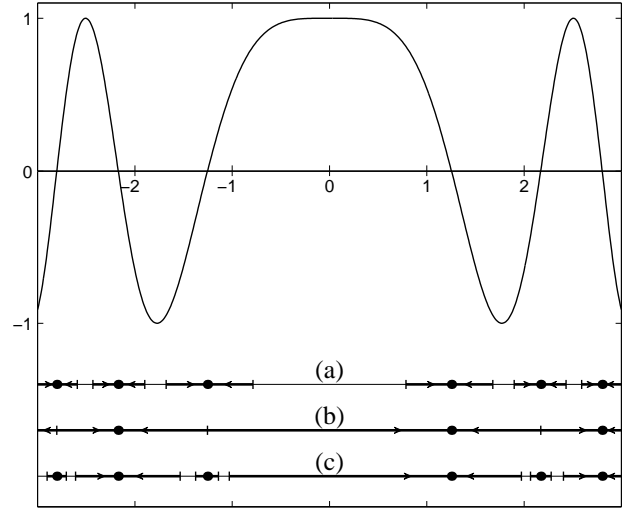}
\vspace*{2mm}
  \caption{ \narrowtext Shown with thick lines are the basins of 
            convergence of (a) the Newton-Raphson (NR) method,
            (b) the Schmelcher-Diakonos (SD) method with 
                $0 < \lambda < 0.3568$ and ${\bf C} = 1$, and 
            (c) our method with $\beta = 4.0$ and ${\bf C} = 1$ 
            to the zeros of a function $\cos(x^2)$ in the 
            interval $(-3, 3)$.  Arrows indicate the direction of
            convergence and large dots are the zeros to which the
            methods converge.} 
  \label{fig:cos}
\end{figure}
\noindent
matrix as in Eq~(\ref{eq:sd}).
In the vicinity of an UPO, the function $g({\bf x})$ tends to zero and 
the NR method is restored.  In fact, it is straightforward to verify 
that the above scheme retains the quadratic convergence.  On the other
hand, away from the solution and for sufficiently large values 
of $\beta$, our scheme is similar to Eq.~(\ref{eq:sd}) 
and thus almost completely preserves the global convergence property 
of the SD method.  This similarity is easily understood, 
since Eq.~(\ref{eq:sd}) is the Euler method with step size 
$\lambda$ for solving the following system of ODEs:
\begin{equation}  \label{eq:odes}
  \frac{{\mathrm d}{\bf x}}{{\mathrm d}s} = 
  {\bf C} {\bf g}({\bf x}).
\end{equation}
On the other hand, Eq.~(\ref{eq:ours}) is the 
{\em semi-implicit} Euler method \cite{PressBook} with step 
size $h = 1/\beta g({\bf x})$ for solving the same system of ODEs.
Consequently, with sufficiently small step size, both methods 
closely follow the solutions to Eq.~(\ref{eq:odes}) and thus share
the global convergence property. 

To illustrate and contrast the convergence properties of the NR,
the SD, and our methods, we consider the following simple example:
finding zeros of the function $g(x) = \cos(x^2)$ in the interval 
$(-3, 3)$.  The basins of convergence for each method are
shown in Fig.~\ref{fig:cos} with the thick arrows.  
The NR method converges to all six zeros and the basins are 
essentially within the linear neighborhood of each 
point \cite{NR_fractal}.  The SD method converges to the solution
$g(x_0) = 0$ if $0 < \lambda < 2/g'(x_0)$ and 
${\bf C} = - {\mathrm sign}(g'(x_0))$.  Diagram (b) in 
Fig.~\ref{fig:cos} shows the basin of convergence for 
$0 < \lambda < 0.3568$ and ${\bf C} = 1$.  Obvious is the global 
character of convergence to zeros
with negative function derivatives,
while zeros with positive derivatives serve as basin boundaries.
With ${\bf C} = -1$ the convergence directions are reversed.
The result of applying our iteration scheme with $\beta = 4.0$ 
and ${\bf C} = 1$ to the same function is shown in the 
diagram (c).  We see that, as in the NR method, all zeros have 
basins of convergence.
However, the basins of zeros with negative function derivatives
cover most of the interval, while the basins of other zeros, as well 
as the intervals between basins, are reduced and become smaller with
increasing value of $\beta$.  Therefore, our scheme combines the
efficiency of the NR method with the global character of 
the SD algorithm. 
\begin{figure}[t]
  \epsfxsize=8.3cm  \epsfbox{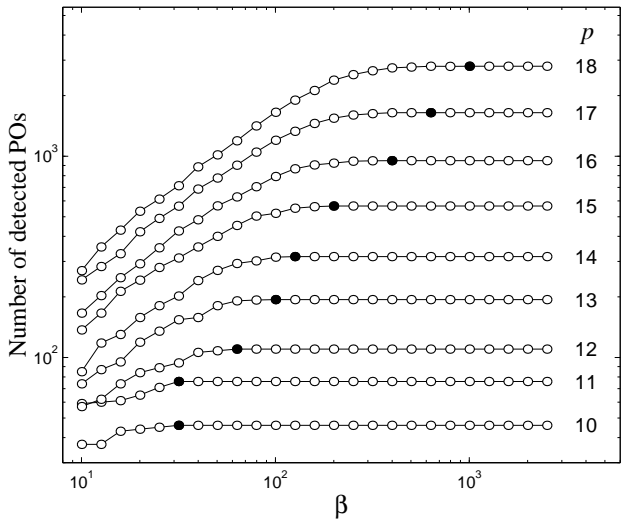}
\vspace*{2mm}
  \caption{Number of detected orbits for different periods in the
           Ikeda-Hammel-Jones-Moloney attractor given in 
           Eq.~(\ref{eq:ikeda}).
           Solid dots indicate the values of $\beta$ above which we 
           always detect a maximum number of UPOs for each period.}
  \label{fig:beta}
\end{figure}

Another important ingredient of our method lies in the selection of
initial points: we find that the most efficient strategy 
for detecting UPOs of period $p$ is to use UPOs of other periods 
as initial points.  This is understandable, since orbit points
cover the attractor in a systematic manner, which reflects the
foliation of the function ${\bf f}^{(p)}({\bf x})$ and its 
iterates.  In cases of the H\'{e}non and the 
Ikeda-Hammel-Jones-Moloney maps,
we are able to detect all UPOs of period $p$ using only orbit points 
of period $p-1$, provided that period $p-1$ orbits exist. 
In more complicated cases of higher-dimensional maps, this simple 
strategy leaves a small fraction of UPOs undetected \cite{Grid}.
However, in all cases, we are able to find these UPOs using period 
$p+1$ points (first we use incomplete set of period $p$ orbits 
to find period $p+1$ points and then use them to complete the 
detection of period $p$ orbits).
The main advantage of using orbit points of neighboring periods as
initial points is that once we establish the strategy for smaller
periods, it works in a similar manner for the detection of
UPOs of large period.  This allows us to claim with confidence
that we detect all UPOs of increasingly longer periods
for general multi-dimensional chaotic maps. 

We now apply our method to detecting UPOs for the following 
Ikeda-Hammel-Jones-Moloney map \cite{Ikeda:1979}:
\begin{eqnarray}   \label{eq:ikeda}
  x' &=& a + b(x\cos{\phi} - y\sin{\phi})\:, \nonumber \\
  y' &=& b(x\sin{\phi} + y\cos{\phi})\:,
\end{eqnarray}
where $\phi = k - \eta/(1 + x^2 + y^2)$, and the parameters
are chosen such that the map has a chaotic attractor:
$a = 1.0$, $b = 0.9$, $k = 0.4$, and $\eta = 6.0$.  Detection of 
UPOs proceeds as follows:  UPOs of period 1 and 2 are quickly
found using several initial points on the attractor.  Starting from
$p = 3$ we use only orbit points of period $p-1$ as initial 
points.  We choose ${\bf C}$ from the set of five matrices
$\{{\bf C}_k | k = 1,\ldots ,5\}$ provided in Ref. \cite{SD:1997},
where ${\bf C}_1 = {\bf 1}$ is the identity matrix.  
The iteration sequence computed from Eq.~(\ref{eq:ours}) is 
terminated when it either converges to an orbit point or leaves 
the chaotic attractor.  The average number of iterations
increases linearly with $\beta$, which is understandable since
$||\delta{\bf x}|| \approx 1/\beta$ for large $\beta$ and away from 
an orbit point.  However, a small fraction of initial points produces 
very long sequences which neither converge to an UPO nor leave the
attractor.  In order to limit the amount of
unproductive computation, we set the maximum number of iterations 
to 4-6 times $\beta$, which is sufficient for the majority of 
iterates to be terminated properly.  The quadratic convergence of our 
scheme allows us to achieve, without much computational effort, 
accuracy limited only by the computer round-off error.  Once the 
sequence converges to an orbit point, we check whether it belongs 
to a yet undetected UPO, and if so, we compute the rest of the 
orbit points by iterating the map and refining the solutions with a 
couple of NR steps [we simply set $\beta = 0$ in 
Eq.~(\ref{eq:ours})\,].

Figure \ref{fig:beta} shows the number of detected UPOs of periods 10
through 18 using different values of $\beta$ in the range from 10 
to 3000.  Note that for every period there exists a value 
$\beta = \beta_{\mathrm min}(p)$ above which we are guaranteed to 
find a maximum number of UPOs.  This feature of our scheme strongly 
suggests that the detected orbits constitute a {\it complete} set 
of UPOs for each period. 
Since $\beta_{\mathrm min}(p)$ is approximately proportional to 
${\mathrm e}^{\alpha p}$, where $\alpha$ is a positive constant,
we can estimate the value of $\beta$ necessary to find all UPOs
of increasingly longer periods.
The numbers of the UPOs for periods up to 13 agree with those of
Schmelcher and Diakonos \cite{SD:1997} except for period 10,
where we have detected one additional orbit. 
The number of orbits of periods 14 through 22, which were not reported
previously, are given in Table~\ref{tab:npos}. 
\begin{table}
  \caption{ Number of distinct UPOs, $n(p)$, and the 
            total number of orbit points or period $p$, $N(p)$,
            for the Ikeda attractor given by
            Eq.~(\ref{eq:ikeda}). Note that $N(p)$ also includes 
            orbit points whose periods are factors of $p$.}
\begin{tabular}{crr}
$p$ & $n(p)$ & $N(p)$ \\ \hline
14 &     317 &   4\,511 \\
15 &     566 &   8\,517 \\
16 &     950 &  15\,327 \\
17 &  1\,646 &  27\,983 \\
18 &  2\,799 &  50\,667 \\
19 &  4\,884 &  92\,797 \\
20 &  8\,404 & 168\,575 \\
21 & 14\,700 & 308\,777 \\
22 & 25\,550 & 562\,939
\end{tabular}
  \label{tab:npos}
\vspace*{-1ex}
\end{table}

\begin{figure}
  \epsfxsize=8.3cm  \epsfbox{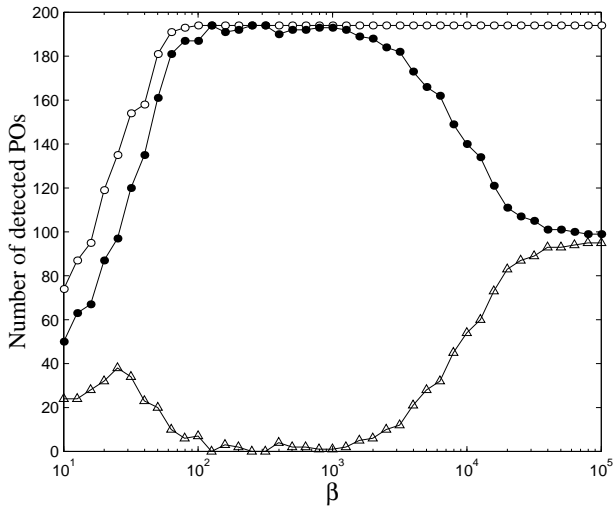}
\vspace*{2mm}
  \caption{Detection of UPOs of period 13 in the 
           Ikeda-Hammel-Jones-Moloney attractor.
           The number of orbits detected with ${\bf C}_1$ is shown
	   with solid dots, while triangles represent the number of 
           additional orbits detected with ${\bf C}_k$, 
           $k = 2,\ldots,5$.  The total number of detected orbits is
           shown with open circles.}
  \label{fig:p13}
\end{figure}

If we monitor the number of orbits detected with different
matrices ${\bf C}$, we note that, for a wide range of values 
of $\beta$, after we use identity matrix ${\bf C}_1$, 
only a few UPOs remain undetected.  For example, 
with $\beta = 5000$ and ${\bf C} = {\bf C}_1$ in Eq.~(\ref{eq:ours}),
our method detects 14\,699 orbits of period 21, and
only one new orbit is detected with ${\bf C} = {\bf C}_2$.
To understand this feature of our method, which is common to all 
the maps tested, we show in 
Fig.~\ref{fig:p13}, for the chaotic attractor in Eq.~(\ref{eq:ikeda}),
the number of period 13 orbits detected with ${\bf C}_1$ (solid dots)
and the number of additional orbits detected with
${\bf C}_k$, $k = 2,\ldots,5$, (triangles).  For $100 < \beta < 1000$,
almost all UPOs are detected with ${\bf C}$ being the identity matrix.
At larger values of $\beta$ the number of thus detected orbits
decreases, but the remaining orbits are always detected with
other matrices.  For $\beta > 10^5$ the numbers converge to those 
of the SD iteration scheme, where about half
of the orbits are detected with ${\bf C}_1$ and the other half
with ${\bf C}_2$ and ${\bf C}_3$.  This behavior of our scheme
follows directly from the convergence considerations of 
Fig.~\ref{fig:cos} and results in a greatly improved efficiency
compared to either the NR or the SD methods.

Finally, we briefly describe the performance of our method for
other maps.  In case of the H\'{e}non map our algorithm works
extremely well and, for the standard parameter values of
$(a, b) = (1.4, 0.3)$, detects all UPOs up to period 29
with $\beta < 500$, ${\bf C} = {\bf C}_1$ and ${\bf C}_2$,  
and using for initialization only orbit points of period $p-1$.
We have also applied our algorithm to detecting UPOs
in the following four-dimensional map: Two coupled Ikeda maps
with coupling in the form:
$\phi_{(1,2)} = k - \eta/(1 + x_{(1,2)}^2 + y_{(1,2)}^2) + 
2\pi\varepsilon(x_{(2,1)} - x_{(1,2)})$, and the parameters are
chosen such that the system has two positive Lyapunov exponents.
We estimate the topological entropy in this system to be
$h_{\mathrm T} \approx 1.6$, and thus the number of orbits grows
extremely fast with increasing orbit length.  We have detected 
complete sets of UPOs up to period 7 with $\beta < 1000$.  
We have found that the reliability of the algorithm was not
affected by the increased dimensionality of the system.
Even though the number of possible matrices ${\bf C}$ in four 
dimensions is 384, only a dozen of them are needed to detect all UPOs.
The necessary set of matrices ${\bf C}$ can be selected empirically
when detecting short UPOs and then used in the detection of longer
orbits. 
   
In conclusion, we have proposed an efficient algorithm for
the detection of UPOs in chaotic systems and
have successfully detected large number of UPOs in several
two- and higher-dimensional maps.  Our method allows for a
verification of the completeness of the detected orbits and
high accuracy limited only by the round-off error.  

This work was supported by AFOSR under Grant No. F49620-98-1-0400 and 
by NSF under Grant No. PHY-9722156.

\ifpreprintsty
\else
\fi


\begin{thebibliography}{10}

\bibitem{UPO_general}
D. Auerbach, P. Cvitanovi\'{c}, J.-P. Eckmann, G. H. Gunaratne,
and I. Procaccia, Phys. Rev. Lett. {\bf 58}, 2387 (1987);
G. H. Gunaratne and I. Procaccia, Phys. Rev. Lett. {\bf 59}, 1377 
(1987); 
D. Auerbach, B. O'Shaughnessy, and I. Procaccia, Phys. Rev. A {\bf 37},
2234 (1988); P. Cvitanovi\'{c} and B. Eckhardt,
Phys. Rev. Lett. {\bf 63}, 823 (1989);
D. Auerbach, Phys. Rev. A {\bf 41}, 6692 (1990);
P. Cvitanovi\'{c}, Focus Issue on Periodic Orbit Theory,
Chaos {\bf 2}, 1992.

\bibitem{GOY:1988}
C. Grebogi, E. Ott, and J. A. Yorke, Phys. Rev. A {\bf 37}, 
1711 (1988);  Y.-C. Lai, Y. Nagai, and C. Grebogi, Phys. Rev. Lett. 
{\bf 79}, 649 (1997).

\bibitem{Gutzwiller:1990}
M.~C. Gutzwiller, {\em Chaos in Classical and Quantum Mechanics} 
(Springer, New York, 1990).

\bibitem{UPO_exp}
D. P. Lathrop and E. J. Kostelich, Phys. Rev. A {\bf 40}, 4028 (1989);
D. Pierson and F. Moss, Phys. Rev. Lett. {\bf 75}, 2124 (1995);
D. Christini and J. J. Collins, Phys. Rev. Lett. {\bf 75}, 2782 (1995);
X. Pei, and F. Moss, Nature (London) {\bf 379}, 619 (1996);
B. Hunt and E. Ott, Phys. Rev. Lett. {\bf 76}, 2254 (1996);
P. So, E. Ott, S. J. Schiff, D. T. Kaplan, T. Sauer, and C. Grebogi,
Phys. Rev. Lett. {\bf 76}, 4705 (1996);
P. So, E. Ott, T. Sauer, B. J. Gluckman, C. Grebogi, and S. J. Schiff,
Phys. Rev. E {\bf 55}, 5398 (1997).

\bibitem{BW:1989}
Particular methods exist for specific systems or in special 
cases, such as the 
method by Biham and Wenzel [O. Biham and W. Wenzel, Phys. Rev. Lett.
{\bf 63}, 819 (1989)] for the H\'{e}non map for which 
a Hamiltonian-like function can be found with extrema located at the
orbit points of UPOs, and the method by Hansen
[K. Hansen, Phys. Rev. E {\bf 52}, 2388 (1995)] which is applicable
to two-dimensional maps if the symbolic dynamics of the map
is known and well ordered.

\bibitem{SD:1997}
P. Schmelcher and F.~K. Diakonos, Phys. Rev. Lett. {\bf 78}, 4733 
(1997); Phys. Rev. E {\bf 57},  2739 (1998).

\bibitem{Ikeda:1979}
K. Ikeda, Opt. Commun. {\bf 30}, 257 (1979);
S. M. Hammel, C. K. R. T. Jones, and J. Moloney, J. Opt. Soc. Am. B
{\bf 2}, 552 (1985).

\bibitem{PressBook}
W.~H. Press, S.~A. Teukolsky, W.~T. Vetterling, and B.~P. Flannery, 
{\em  Numerical Recipes in Fortran}, 2nd ed. (Cambridge University 
Press, Cambridge, 1992).

\bibitem{Impractical}
It is practically impossible to use the SD method to detect
complete sets of UPOs for periods above 20 because the amount of
computation required grows exponentially at a much higher rate than
that of our method.

\bibitem{DSB:1998}
F.~K. Diakonos, P. Schmelcher, and O. Biham, Phys. Rev. Lett. 
{\bf 81},  4349  (1998).

\bibitem{Convergence}
Indeed, close to a zero point, 
the corrections from Eq.~(\ref{eq:sd}) are proportional to the
deviation (linear convergence), while corrections determined from
the NR method yield an error that is proportional to the 
square of the deviation (quadratic convergence).
 
\bibitem{NR_fractal}
Even though the Newton-Raphson method generally has 
a fractal basin structure, we show only intervals adjacent to the 
solution, as they are the most reliable source of starting points. 

\bibitem{Grid}  
For relatively short orbits we verify the
completeness of the detected sets by initializing our iteration scheme
on a fine grid of initial points.

\end{thebibliography}
\end{document}